# An institutional study on plan quality and variation of manual forward planning for Gamma Knife radiosurgery for vestibular schwannoma

**Zhen Tian, Tonghe Wang, Xiaofeng Yang, Matt D. Giles, Elizabeth Butker, Matthew C. Walb, Tian Liu, Shannon Kahn**

Department of Radiation Oncology, Emory University, Atlanta, GA30022

Emails: zhen.tian@emory.edu

**Abstract:** Due to the complexity and cumbersomeness of Gamma Knife (GK) manual forward planning, the quality of the resulting treatment plans heavily depends on the planners' skill, experience and the amount of effort devoted to plan development. Hence, GK plan quality may vary significantly among institutions and planners, and even for a same planner at different cases. This is particularly a concern for challenging cases with complicated geometry, such as vestibular schwannoma cases. The purpose of this retrospective study is to investigate the plan quality and variation in the manually forward planned, clinically acceptable GK treatment plans of 22 previous vestibular schwannoma cases. Considering the impacts of different patient geometry and different trade-offs among the planning objectives in GK planning, it is difficult to objectively assess the plan quality across different cases. To reduce these confounding factors on plan quality assessment, we employed our recently developed multiresolution-level inverse planning algorithm to generate a golden plan for each case, which is expected to be on or close to the pareto surface with a similar trade-off as used in the manual plan. The plan quality of the manual plan is then quantified in terms of its deviation from the golden plan. A scoring criterion between 0-100 was designed to calculate a final score for each manual plan to simplify our analysis. Large quality variation was observed in these 22 cases, with two cases having a score lower than 75, three cases scoring between 80 and 85, two cases between 85 and 90, eight cases between 90 and 95, and seven cases higher than 95. Inter- and intra- planner variability was also observed in our study. This large variation in GK manual planning deserves high attention, and merits further investigation on how to reduce the variation in GK treatment plan quality.

# 1    INTRODUCTION

Gamma Knife (GK) radiosurgery has emerged as an important and safe alternative to traditional neurosurgery for vestibular schwannoma[1-3]. The most advanced GK treatment units, i.e. Perfexion™ and Icon™ (Elekta Instrument AB Stockholm, Stockholm, Sweden), focus 192 narrow $^{60}$Co beams on a single point. This design can eradicate tumor viability with submillimeter accuracy, while providing a very steep dose gradient to minimize damage to surrounding normal tissues in order to reduce risk of side effects and improve cochlear sparing.

Manual forward planning is the current standard GK planning method, in which planners adjust the number and locations of the irradiation focal points, beam sizes, shapes and beam-on time in a trial-and-error schema to achieve a clinically acceptable treatment plan. Ideally, any location within the target volume is a potential beam isocenter. In addition, since the 192 sources are arranged on eight individual sectors that can be driven independently to one of four different states (i.e., 4, 8 or 16 mm collimator sizes or be blocked), there are 65,536 (i.e. $4^8$) available beam shapes to be used at each isocenter. With such large degrees of freedom involved in planning, manual forward planning is challenging and cumbersome, particularly for vestibular schwannoma cases due to the irregular-shaped target and its proximity to brainstem and cochlea[4-9]. It is impossible for planners to explore the whole solution space via manual forward planning to develop the best plan for each specific patient. The resulting plan quality hence heavily relies on planners' skills, experiences, and available time to develop the plan, and can vary significantly from case to case, planner to planner and institution to institution.

Although there have been studies comparing plan qualities between GK, CyberKnife (Accuray Incorporated company, Sunnyvale, CA) and linac-based radiosurgery[10-18], there are few studies which focus on the plan quality and variation of GK radiosurgery itself. In this study, we aim to perform a retrospective study to investigate the plan quality and variation of our previous treated GK treatment plans that have been generated via manual forward planning for our previous vestibular schwannoma cases. There are two challenges that are encountered when attempting to assess the plan quality and variation of GK radiosurgery. First, the best plan quality that can be achieved for each patient is essentially determined by his or her specific geometry. For instance, due to the nature of GK radiosurgery, a target with an irregular shape is prone to inferior dose conformity compared to a target with a spherical shape. Hence, an equivalent dose conformity index achieved in a treatment plan doesn't necessarily translate to the same plan quality for different patients. As previously mentioned, the large number of variables involved in GK planning makes it impossible for planners to explore the whole solution space via manual forward planning. Thus, we do not have a priori knowledge of the best plan quality that can be achieved for each specific patient geometry, which makes it difficult to quantify and compare the plan quality. Secondly, a trade-off among multiple planning objectives is often needed when all the planning objectives cannot be met at the same time. For instance, when a critical organ is very close to the target, physicians need to determine the priority between target coverage and organ sparing based on the health conditions of the particular patient. Another trade-off we often encounter in GK planning is between dose conformity and beam-on time. High dose conformity is often associated with small collimator sizes which are more flexible for beam shaping but prolongs the beam-on time. Different trade-offs adopted on the same case would result in different plan properties, which doesn't necessarily mean different plan qualities.

We have recently developed a multi-resolution-level (MRL) inverse planning approach for GK

radiosurgery, which provides a feasible solution to approach the intractable large-scale global optimization problem iteratively, and yields a pareto optimal plan or a plan close to the pareto surface[19]. In this study, we employed this MRL inverse planning approach to conquer the two aforementioned challenges to enable our investigation of the plan quality and variation of GK manual forward planning for the previous vestibular schwannoma treatment in our institution. In addition, a scoring mechanism was designed to calculate a final score for each manual plan to simplify our analysis.

## 2 METHODS AND MATERIALS

*2.1 Patient cases*

With IRB approval, we retrospectively reviewed 21 patients treated for vestibular schwannoma using GK radiosurgery with a Leksell GK ICON™ treatment unit at our institution from 2017-2019. All patients received a single fraction of 12.5 Gy, prescribed to 50% isodose level to allow higher dose within the tumor itself. For each patient, the treatment plan was developed by our GK medical physicists using Leksell Gamma Plan (LGP) treatment planning system (Elekta Instrument AB Stockholm, Stockholm, Sweden) via manual forward planning, and reviewed and approved by radiation oncologists and neurosurgeons for treatment. Several rounds of consultation between planners and physicians often occurred during treatment planning to achieve a satisfying plan of physicians' preferred trade-off for each patient. The related plan information of the original manual plans for these 22 cases are listed in Table 1 in reverse chronological order.

Table 1. Plan information of the original manual plans for our previous 21 vestibular schwannoma cases treated with GK radiosurgery with a Leksell GK ICON™ unit.

| case | Target volume (cc) | Planner | Coverage | Selectivity | Gradient index | CI50* | $D_{BS,0.1cc}$ (Gy) | $D_{co,mean}$ (Gy) | BOT* (min) |
|---|---|---|---|---|---|---|---|---|---|
| 1 | 0.193 | P2 | 1.00 | 0.56 | 3.10 | 5.54 | 1.1 | 4.7 | 32.55 |
| 2 | 0.280 | P2 | 1.00 | 0.58 | 2.98 | 5.14 | 2.4 | 3.6 | 27.56 |
| 3 | 2.273 | P2 | 1.00 | 0.73 | 2.65 | 3.63 | 11.7 | 10.7 | 60.96 |
| 4 | 1.217 | P1 | 1.00 | 0.71 | 2.83 | 3.99 | 1.2 | 0.1 | 38.22 |
| 5 | 0.777 | P1 | 1.00 | 0.69 | 2.77 | 4.01 | 2.8 | 4.2 | 60.97 |
| 6 | 0.303 | P2 | 1.00 | 0.71 | 2.97 | 4.18 | 2.0 | 6.9 | 31.91 |
| 7 | 0.79 | P1 | 1.00 | 0.65 | 2.91 | 4.48 | 7.6 | 8.8 | 45.95 |
| 8 | 0.289 | P1 | 1.00 | 0.56 | 2.96 | 5.29 | 2.5 | 3.7 | 41.24 |
| 9 | 1.4006 | P2 | 1.00 | 0.63 | 2.72 | 4.32 | 8.7 | 6.5 | 51.10 |
| 10 | 0.093 | P2 | 1.00 | 0.44 | 3.48 | 7.91 | 1.4 | 8.1 | 27.95 |
| 11 | 0.212 | P1 | 1.00 | 0.45 | 3.15 | 7.00 | 1.3 | 4.9 | 51.78 |
| 12 | 2.779 | P2 | 1.00 | 0.81 | 2.67 | 3.30 | 10.0 | 4.0 | 42.33 |
| 13 | 0.379 | P2 | 1.00 | 0.60 | 2.86 | 4.77 | 2.3 | 7.5 | 20.88 |
| 14 | 2.093 | P2 | 1.00 | 0.77 | 2.84 | 3.69 | 9.5 | 6.4 | 32.60 |
| 15 | 0.284 | P3 | 0.99 | 0.51 | 2.94 | 5.76 | 1.3 | 4.0 | 22.40 |
| 16 | 0.283 | P1 | 0.99 | 0.43 | 3.03 | 7.05 | 1.8 | 4.0 | 42.41 |
| 17 | 0.874 | P1 | 1.00 | 0.70 | 2.84 | 4.06 | 5.9 | 4.0 | 32.13 |
| 18 | 1.311 | P1 | 1.00 | 0.78 | 2.61 | 3.35 | 10.1 | 3.0 | 34.51 |
| 19 | 2.13 | P3 | 1.00 | 0.63 | 3.27 | 5.19 | 4.3 | 2.4 | 34.68 |
| 20 | 0.313 | P3 | 1.00 | 0.50 | 2.97 | 5.94 | 5.1 | 4.2 | 26.19 |
| 21 | 0.418 | P1 | 1.00 | 0.65 | 3.03 | 4.66 | 2.5 | 4.9 | 36.52 |
| 22 | 0.415 | P1 | 1.00 | 0.67 | 3.17 | 4.73 | 5.5 | 4.0 | 39.70 |

\* BOT listed here is the total beam-on time normalized at the initial dose rate 3.534 Gy/min;
\*CI50 denotes conformity index at 50% isodose level as defined in Eq.(4).

*2.2 Study design*

Our MRL inverse planning approach, which will be briefly introduced in subsection 2.4, was used to re-plan these previous cases. The same planning guidelines used in our institution to guide manual forward planning for GK radiosurgery of vestibular schwannoma were adopted to guide our re-planning as well: (1) 100% of prescription dose must be received by at least 99% of the target volume, that is, target coverage ≥ 0.99; (2) The maximum dose to 0.1 cc of brainstem must not exceed 12 Gy, that is, $D_{BS,0.1cc} \leq 12$ Gy; (3) Try to keep the mean dose of the ipsilateral cochlea below 4 Gy, that is, $D_{co,mean} \leq 4$ Gy; (4) Try to maximize selectivity and minimize gradient index to spare the nearby normal tissues as much as possible. During our inverse planning, we tried to realize similar trade-off that was adopted in the original manual plan for fair comparison. The obtained plans were expected to be pareto optimal plans or plans close to the pareto surface, and hence were considered as "golden" plan to help us to reduce the impact of different patient geometry on plan quality evaluation.

GK plan is usually evaluated in terms of coverage (as defined in Eq.(1)), selectivity (Eq.(2)), gradient index (Eq.(3)), OAR dose, and total beam-on time.

$$\text{coverage} = \frac{TV \cap PIV}{TV}. \tag{1}$$

$$\text{selectivity} = \frac{TV \cap PIV}{PIV}, \tag{2}$$

$$\text{gradient index} = \frac{PIV_{R_x/2}}{PIV}. \tag{3}$$

Here, TV and PIV represent the target volume and the planning isodose volume, respectively. $PIV_{R_x/2}$ represents the volume that receives at least half of the prescription dose. Gradient index was proposed to compare the dose gradient outside the target for the treatment plans of equal dose conformity. It cannot be directly used to compare plans of different conformity level. Hence, instead of gradient index, we used conformity index at 50% isodose line (CI50, as defined in Eq.(4)) in this study to access the intermediate dose spillage for each treatment plan.

$$\text{CI50} = \frac{PIV_{R_x/2}}{TV}. \tag{4}$$

The values of these metric are affected by the specific geometry of each patient as well as the trade-offs preferred by the physicians, and therefore cannot be directly compared between different cases. Therefore, in this study we calculated the deviation of the original manual plans from the golden plans obtained via MRL inverse planning with a similar trade-off. A scoring mechanism was also proposed in this study to calculate a single score for each case. Our scoring mechanism has six score components, corresponding to coverage, selectivity, CI50, $D_{BS,0.1cc}$, $D_{co,mean}$, and total beam-on time. Their values are set to be 100 for the golden plan, while the values of these components for the manual plan are scored according to the scoring criterion listed in Table 2. The minimum values of these score components are set to be 0, and the values are allowed to exceed 100 if a particular plan metric of the manual plan is better than that of the golden plan. Specifically, the coverage score, denoted as $S_{coverage}$, is decreased by 10 per 0.01 coverage loss of the manual plan compared to the golden plan. To comply with our aforementioned planning guidelines, the value of $S_{coverage}$ becomes 0 if the coverage is lower than 0.99. The selectivity score of the manual plan was decreased by 3.3 per 0.0275 reduction of selectivity. Similarly, the CI50 score was decreased by 3.3 per 0.09 increase in CI50, and the BOT score was decreased by 3.3 per 1.8% increase of the total beam-on time. These scoring criteria were made based on the observations in our experiments, in which we changed the prescribed isodose line of the golden plans to 1% higher without adjusting anything else of the plans. It was observed that the coverage deteriorated from 1 to 0.99. At the same time, the selectivity was increased by 0.0275 averagely, CI50 was improved by 0.09, and the average reduction of BOT was ~1.8%. The resulting increase of $D_{BS,0.1cc}$ and $D_{co,mean}$ were found to be negligible. The aforementioned scoring criteria were made to yield the same score for two sets of the golden plans, considering they are of same quality level but different trade-offs. The scoring creteria for $D_{BS,0.1cc}$ and $D_{co,mean}$ were made emperically based on our institutional planning guidelines and our GK planning experiences. Specifically, the score for $D_{BS,0.1cc}$ is decreased by 5.0 per 1 Gy increase compared to the golden plan, and becomes 0 if $D_{BS,0.1cc}$ exceeds our 12Gy tolerance. The score for $D_{co,mean}$ is decreased by 5.0 per 1 Gy increase when $D_{co,mean}$ is within our desired 4Gy tolerance, and decreased by 10 per 1 Gy increase when $D_{co,mean}$ exceeds it. This scoring criterion does not purport to describe the best clinical plan, but serves here as a relatively objective method to quantify the plan quality disparity between the manual plan and the golden plan for each specific patient case in this study. The standard deviation of these scores was calculated to assess the plan quality variation among all the cases. The plan quality variation was also assessed for each planner.

**Table 2**. Scoring mechanism proposed in this study to calculate a single score for each case in terms of the quality deviation of the original manual plans from the golden plans.

| |
|---|
| $S_{coverage} = \begin{cases} 100 + \dfrac{coverage^{manual} - coverage^{golden}}{0.01} \times 10, if\, coverage^{manual} \geq 0.99 \\ 0, if\, coverage^{manual} < 0.99 \end{cases}$ |
| $S_{selectivity} = \max(100 + \dfrac{selectivity^{manual} - selectivity^{golden}}{0.0275} \times 3.3, 0)$ |
| $S_{CI50} = \max(100 - \dfrac{CI50^{manual} - CI50^{golden}}{0.09} \times 3.3, 0)$ |
| $S_{BOT} = \max(100 - \dfrac{(BOT_{manual} - BOT_{golden})/BOT_{golden}}{0.018} \times 3.3, 0)$ |
| $S_{D_{BS,0.1cc}} = \begin{cases} 100 - \dfrac{D_{BS,0.1cc}^{manual} - D_{BS,0.1cc}^{golden}}{1.0\, Gy} \times 5, if\, D_{BS,0.1cc}^{manual} \leq 12Gy \\ 0, if\, D_{BS,0.1cc}^{manual} > 12Gy \end{cases}$ |
| $S_{D_{co,mean}} = \begin{cases} 100 - \dfrac{D_{co,mean}^{manual} - D_{co,mean}^{golden}}{1.0\, Gy} \times 5, if\, D_{co,mean}^{manual}\, and\, D_{co,mean}^{golden} \leq 4Gy \\ 100 - \dfrac{D_{co,mean}^{manual} - D_{co,mean}^{golden}}{1.0\, Gy} \times 10, if\, D_{co,mean}^{manual}\, and\, D_{co,mean}^{golden} > 4Gy \\ 100 - \dfrac{D_{co,mean}^{manual} - 4}{1.0\, Gy} \times 10 - \dfrac{4 - D_{co,mean}^{golden}}{1.0\, Gy} \times 5, if\, D_{co,mean}^{manual} > 4Gy\, and\, D_{co,mean}^{golden} \leq 4Gy \\ 100 - \dfrac{D_{co,mean}^{manual} - 4}{1.0\, Gy} \times 5 - \dfrac{4 - D_{co,mean}^{golden}}{1.0\, Gy} \times 10, if\, D_{co,mean}^{manual} \leq 4Gy\, and\, D_{co,mean}^{golden} > 4Gy \end{cases}$ |
| $S_{total} = (S_{coverage} + S_{selectivity} + S_{CI50} + S_{BOT} + S_{D_{BS,0.1cc}} + S_{D_{co,mean}})/6$ |

*2.3 MRL inverse planning*

MRL inverse planning approach is an iterative approach that we recently developed to solve the intractable large-scale GK global optimization problem to find a global optimal plan or an approximate global optimal plan[19]. In this approach, several rounds of optimization are performed with a progressively increased resolution used for isocenter candidates. Specifically, at the beginning a coarse 3D grid is employed to discretize all the locations within the target into grid points and used as initial isocenter candidates. A convex optimization problem was solved to optimize the beam-on time for each sector of each available collimator size at each isocenter candidate. The optimal isocenters are the isocenter candidates that obtain non-zero beam-on time during the optimization. In this way, the isocenter location is actually optimized along with the collimator sizes and beam durations, unlike the other sequential inverse planning algorithms which pre-determine the isocenters based on some geometric methods and only optimize the beam shape and duration at these predetermined isocenters. At the next round of optimization, these optimal isocenters determined at the previous resolution level are used as new isocenter candidates. A finer grid resolution is also used to add the neighboring grid points of these optimal isocenters into the isocenter candidate set to fine tune the isocenter locations and search for a better solution if it exists. The objective function that is used at each round of optimization but with different sets of isocenter candidates is formulated as

$$\begin{aligned}\text{minimize}_t \ & \frac{\omega_{TH}}{N_T} \sum_{i \in I_T} \max(D_i - D_{TH}, 0) + \frac{\omega_{TL}}{N_T} \sum_{i \in I_T} \max(D_{TL} - D_i, 0) \\ & + \frac{\omega_{IS}}{N_{IS}} \sum_{i \in I_{IS}} \max(D_i - D_{IS}, 0) + \frac{\omega_{OS}}{N_{OS}} \sum_{i \in I_{OS}} \max(D_i - D_{OS}, 0) \\ & + \frac{\omega_{BS}}{N_{BS}} \sum_{i \in BS} max(D_i - D_{BS}) + \frac{\omega_{co}}{N_{co}} \sum_{i \in I_{co}} (D_i - D_{co}) \\ & + \omega_{BOT} \sum_{n=1}^{N} \max_{m=1,2,\ldots,8} \left( \sum_{c=1}^{3} t(c, m, n) \right), \end{aligned} \quad (5)$$

subject $to \ t \geq 0$.

Here, $t$ is the beam-on time to be optimized for each sector candidate, that is, each physical sector ($m = 1, 2, \ldots, 8$) with a collimator size ($c = 1, 2, 3$ corresponding to 4 mm, 8 mm and 16 mm collimators) at an isocenter candidate ($n = 1, 2, \ldots, N_{isoc}$). Seven planning objectives are used in this objective function, corresponding to target maximum and minimum doses, maximum doses received by the inner and outer shells of the target, brainstem maximum dose, mean dose of the ipsilateral cochlea and total beam-on time, respectively. $\omega_X$ denotes the user-specified priority for the corresponding objective, which were manually adjusted in our study to achieve the similar trade-offs adopted in the original manual plans. After the optimization, sectors with non-zero beam-on time at a same isocenter are grouped into composite shots for delivery via shot sequencing. In this study, the obtained golden plans were put back into the LGP system for final dose calculation.

## 3 RESULTS

The plan metrics of the golden plans obtained via MRL inverse planning are listed in Table 3, and compared to the corresponding values of the forward planned manual plans originally shown in Table 2. The difference of these metric values between the manual plans and the golden plans are calculated and listed in Table 4. The scores of the manual plans in terms of their deviation from the corresponding golden plan were calculated using our scoring mechanism and listed in Table 5. These scores have been reviewed by our physicians and considered to be reasonable values.

**Table 3**. Plan metrics of the original manual plans for our previous cases, and those of the golden plans obtained by MRL inverse planning for each case for plan quality evaluation purpose.

| case | Original manual plans | | | | | | | Golden plans obtained by MRL inverse planning | | | | | | |
|---|---|---|---|---|---|---|---|---|---|---|---|---|---|---|
| | Coverage | Selectivity | Gradient index | CI50 | $D_{BS,0.1cc}$ (Gy) | $D_{co,mean}$ (Gy) | BOT* (min) | Coverage | Selectivity | Gradient index | CI50 | $D_{BS,0.1cc}$ (Gy) | $D_{co,mean}$ (Gy) | BOT* (min) |
| 1 | 1.00 | 0.56 | 3.10 | 5.54 | 1.1 | 4.7 | 32.55 | 1.00 | 0.65 | 3.61 | 5.55 | 1.1 | 4.7 | 27.66 |
| 2 | 1.00 | 0.58 | 2.98 | 5.14 | 2.4 | 3.6 | 27.56 | 1.00 | 0.74 | 3.26 | 4.41 | 2.1 | 3.8 | 26.69 |
| 3 | 1.00 | 0.73 | 2.65 | 3.63 | 11.7 | 10.7 | 60.96 | 1.00 | 0.76 | 2.79 | 3.67 | 11.6 | 9.9 | 41.98 |
| 4 | 1.00 | 0.71 | 2.83 | 3.99 | 1.2 | 0.1 | 38.22 | 1.00 | 0.79 | 3.12 | 3.95 | 1.3 | 0.1 | 27.05 |
| 5 | 1.00 | 0.69 | 2.77 | 4.01 | 2.8 | 4.2 | 60.97 | 1.00 | 0.82 | 3.02 | 3.68 | 2.9 | 3.6 | 41.29 |
| 6 | 1.00 | 0.71 | 2.97 | 4.18 | 2.0 | 6.9 | 31.91 | 1.00 | 0.73 | 2.83 | 3.88 | 2.3 | 6.6 | 34.23 |
| 7 | 1.00 | 0.65 | 2.91 | 4.48 | 7.6 | 8.8 | 45.95 | 1.00 | 0.71 | 2.80 | 3.94 | 7.6 | 8.0 | 43.02 |
| 8 | 1.00 | 0.56 | 2.96 | 5.29 | 2.5 | 3.7 | 41.24 | 1.00 | 0.69 | 3.48 | 5.04 | 3.4 | 3.6 | 29.75 |
| 9 | 1.00 | 0.63 | 2.72 | 4.32 | 8.7 | 6.5 | 51.10 | 1.00 | 0.71 | 2.75 | 3.87 | 8.1 | 5.9 | 50.50 |
| 10 | 1.00 | 0.44 | 3.48 | 7.91 | 1.4 | 8.1 | 27.95 | 1.00 | 0.55 | 3.32 | 6.04 | 1.2 | 8.3 | 27.47 |
| 11 | 1.00 | 0.45 | 3.15 | 7.00 | 1.3 | 4.9 | 51.78 | 1.00 | 0.73 | 3.07 | 4.21 | 1.2 | 4.6 | 45.08 |
| 12 | 1.00 | 0.81 | 2.67 | 3.30 | 10.0 | 4.0 | 42.33 | 1.00 | 0.87 | 2.71 | 3.11 | 10.2 | 3.0 | 37.06 |
| 13 | 1.00 | 0.60 | 2.86 | 4.77 | 2.3 | 7.5 | 20.88 | 1.00 | 0.71 | 3.20 | 4.51 | 2.2 | 7.0 | 20.88 |
| 14 | 1.00 | 0.77 | 2.84 | 3.69 | 9.5 | 6.4 | 32.60 | 1.00 | 0.76 | 2.67 | 3.51 | 10.7 | 5.5 | 32.98 |
| 15 | 0.99 | 0.51 | 2.94 | 5.76 | 1.3 | 4.0 | 22.40 | 0.99 | 0.70 | 3.71 | 5.30 | 1.8 | 3.4 | 23.74 |
| 16 | 0.99 | 0.43 | 3.03 | 7.05 | 1.8 | 4.0 | 42.41 | 1.00 | 0.69 | 3.35 | 4.86 | 1.9 | 3.6 | 36.49 |

| 17 | 1.00 | 0.70 | 2.84 | 4.06 | 5.9 | 4.0 | 32.13 | 1.00 | 0.77 | 3.29 | 4.27 | 5.7 | 3.8 | 30.44 |
| 18 | 1.00 | 0.78 | 2.61 | 3.35 | 10.1 | 3.0 | 34.51 | 1.00 | 0.87 | 2.92 | 3.36 | 9.9 | 2.3 | 32.78 |
| 19 | 1.00 | 0.63 | 3.27 | 5.19 | 4.3 | 2.4 | 34.68 | 1.00 | 0.67 | 2.91 | 4.34 | 4.5 | 1.5 | 35.21 |
| 20 | 1.00 | 0.50 | 2.97 | 5.94 | 5.1 | 4.2 | 26.19 | 1.00 | 0.56 | 2.87 | 5.13 | 4.3 | 3.7 | 25.90 |
| 21 | 1.00 | 0.65 | 3.03 | 4.66 | 2.5 | 4.9 | 36.52 | 1.00 | 0.65 | 2.79 | 4.29 | 2.3 | 4.7 | 36.03 |
| 22 | 1.00 | 0.67 | 3.17 | 4.73 | 5.5 | 4.0 | 39.70 | 1.00 | 0.67 | 3.05 | 4.55 | 6.8 | 3.5 | 38.71 |

\* BOT listed here is the total beam-on time normalized at the initial dose rate 3.534 Gy/min

The best plan quality that can be achieved for each patient depends on their specific geometry, while different trade-offs adopted in planning result in different values of plan metrics. Therefore, it is difficult to quantify the plan quality for each case and the quality variation among different patients. For instance, case 4 and case 6 have different geometries in terms of target volume, shape, and the distance between target to brainstem and cochlea, as shown in the third and fifth columns in Fig. 1. The original manual plans for case 4 and case 6 achieved the same target coverage (1.0) and selectivity (0.71), and similar CI50 (3.99 and 4.18, respectively). The manual plan of case 6 has slightly higher brainstem 0.1cc dose than that of case 4 (2.0 Gy vs. 1.2 Gy) and much higher cochlea mean dose (6.9 Gy vs. 0.1 Gy), which is likely due to the relatively smaller distance between the target target and these two OARs. The BOT of the manual plan for case 6 is shorter than that of case 4 (31.91 min vs. 38.22 min), which may be due to its relatively smaller target volume (0.303 cc vs. 1.217 cc) and/or the different trade-off between cochlea mean dose and BOT adopted for this case. Without clearly knowing how much the different geometry affect each metric given a specified trade-off, it is difficult to judge whether the plan quality of the manual plan of case 6 is comparable, better, or worse than that of case 4 when comparing these two plans directly, although both plans have been reviewed and approved by physicians for treatment. In contrast, when comparing the manual plan with the corresponding golden plan obtained with similar trade-off, the distance of the manual plan from the pareto surface can be quantified to assess its quality level. For case 4, with the same coverage and similar CI50 and OAR doses being obtained, the golden plan has higher selectivity (0.79 vs. 0.71) and much shorter BOT (27.05 min vs. 38.22 min) compared to the manual plan. For case 6, the metrics of its golden plan are close to those of the manual plan, i.e., same coverage, slightly higher selectivity (0.73 vs. 0.71) and lower CI50 (3.88 vs. 4.18), 0.3 Gy higher in brainstem dose and 0.3 Gy lower in cochlea dose, and 2.32 min longer of BOT. It can be concluded that the distance from the manual plan to the pareto surface is larger in case 4 than that in case 6, hence the manual plan of case 4 is of worse quality than case 6, although their plan metrics are of comparable values. This conclusion is also reflected via the final scores. The manual plan of case 4 got a final score of 85.5 mainly due to the low BOT score (23.5), while the manual plan of case 6 got a final score of 99.6.

**Table 4**. Deviations of the manual plans from the golden plans

| case | $\Delta_{coverage}$ | $\Delta_{selectivity}$ | $\Delta_{CI50}$ | $\Delta_{D_{BS,0.01cc}}$ (Gy) | $\Delta_{D_{co,mean}}$ (Gy) | $\Delta_{BOT}$ (min) | $\Delta_{BOT}$ (%) |
|---|---|---|---|---|---|---|---|
| 1 | 0 | -0.09 | -0.01 | 0 | 0 | 4.89 | 17.68 |
| 2 | 0 | -0.16 | 0.73 | 0.30 | -0.20 | 0.87 | 3.26 |
| 3 | 0 | -0.03 | -0.04 | 0.10 | 0.80 | 18.98 | 45.21 |
| 4 | 0 | -0.08 | 0.04 | -0.10 | 0 | 11.17 | 41.29 |
| 5 | 0 | -0.13 | 0.33 | -0.10 | 0.60 | 19.68 | 47.66 |
| 6 | 0 | -0.02 | 0.30 | -0.30 | 0.30 | -2.32 | -6.78 |
| 7 | 0 | -0.06 | 0.54 | 0 | 0.80 | 2.93 | 6.81 |
| 8 | 0 | -0.13 | 0.25 | -0.90 | 0.10 | 11.49 | 38.62 |
| 9 | 0 | -0.08 | 0.45 | 0.60 | 0.60 | 0.60 | 1.19 |
| 10 | 0 | -0.11 | 1.87 | 0.20 | -0.20 | 0.48 | 1.75 |
| 11 | 0 | -0.28 | 2.79 | 0.10 | 0.30 | 6.70 | 14.86 |
| 12 | 0 | -0.06 | 0.19 | -0.20 | 1.00 | 5.27 | 14.22 |
| 13 | 0 | -0.11 | 0.26 | 0.10 | 0.50 | 0.00 | 0.00 |
| 14 | 0 | 0.01 | 0.18 | -1.20 | 0.90 | -0.38 | -1.15 |
| 15 | 0 | -0.19 | 0.46 | -0.50 | 0.60 | -1.34 | -5.64 |
| 16 | -0.01 | -0.26 | 2.19 | -0.10 | 0.40 | 5.92 | 16.22 |
| 17 | 0 | -0.07 | -0.21 | 0.20 | 0.20 | 1.69 | 5.55 |
| 18 | 0 | -0.09 | -0.01 | 0.20 | 0.70 | 1.73 | 5.28 |
| 19 | 0 | -0.04 | 0.85 | -0.20 | 0.90 | -0.53 | -1.51 |
| 20 | 0 | -0.06 | 0.81 | 0.80 | 0.50 | 0.29 | 1.12 |
| 21 | 0 | 0 | 0.37 | 0.20 | 0.20 | 0.49 | 1.36 |
| 22 | 0 | 0 | 0.18 | -1.30 | 0.50 | 0.99 | 2.56 |
| mean | -0.0004 | -0.09 | 0.53 | -0.10 | 0.41 | 3.90 | 10.88 |
| SD | 0.002 | 0.07 | 0.77 | 0.50 | 0.36 | 6.07 | 16.51 |

It can be observed from Table 4 that the deviation of the original manual plan from the golden plan varies for difference cases. Specifically, the deviation ranges from -0.01 to 0 in coverage, -0.28 to 0.01 in selectivity, -0.21 to 2.79 in CI50, -1.3 Gy to 0.8 Gy in brainstem 0.1cc dose, -0.2 Gy to 1.0 Gy in cochlea mean dose, and -2.32 min to 19.68 min in BOT (which is -6.78% to 47.66% of BOT of the golden plans). Averaged over all the cases, the mean deviation ± standard deviation is -0.0004 ± 0.002 in coverage, -0.09 ± 0.07 in selectivity, 0.53 ± 0.77 in CI50, -0.1 ± 0.5 Gy in brainstem dose, 0.41 ± 0.36 Gy in cochlea dose, and 3.90 ± 6.07 min in BOT (i.e., 10.88% ± 16.51% of BOT of the golden plans). Correspondingly, the final scores of these manual plans range from 72.5 to 99.6, with two cases having a score lower than 75, three cases scoring between 80 and 85, two case between 85 and 90, eight cases between 90 and 95, and seven cases higher than 95. The mean score of these manual plans is 90.6 ± 7.7.

The two cases of the lowest scores (i.e., cases 11 and 16 with a score of 72.5 and 74.3, respectively) and the two cases of the highest scores (i.e., cases 6 and 14 with a score of 99.6 and 98.9, respectively) are shown in Figure 1. It can be seen that the manual plans of cases 11 and 16 have much worse dose conformity than their golden plans, while the manual plans of cases 6 and 14 have comparable dose conformity relative to their golden plans. In cases 10 and 15, the BOT of the manual plans are 14.86% and 15.22% longer than that of their golden plans. In contrast, the BOT of the manual plans in cases 6 and 14 are 6.78% and 5.64% shorter than their golden plans.

**Table 5**. Scores calculated for the manual plans using our scoring mechanism

| case | planner | $S_{coverage}$ | $S_{selectivity}$ | $S_{CI50}$ | $S_{D_{BS,0.01cc}}$ | $S_{D_{co,mean}}$ | $S_{BOT}$ | $S_{total}$ |
|---|---|---|---|---|---|---|---|---|
| 1 | P2 | 100 | 89.1 | 100.4 | 100 | 100 | 67.3 | 92.8 |
| 2 | P2 | 100 | 80.6 | 73.0 | 98.5 | 101 | 94.0 | 91.2 |
| 3 | P2 | 100 | 96.4 | 101.5 | 99.5 | 92.0 | 16.3 | 84.3 |
| 4 | P1 | 100 | 90.3 | 98.5 | 100.5 | 100.0 | 23.5 | 85.5 |
| 5 | P1 | 100 | 84.2 | 87.8 | 100.5 | 96.0 | 11.7 | 80.0 |
| 6 | P2 | 100 | 97.6 | 88.9 | 101.5 | 97.0 | 112.6 | 99.6 |
| 7 | P1 | 100 | 92.7 | 80.0 | 100 | 92.0 | 87.4 | 92.0 |
| 8 | P1 | 100 | 84.2 | 90.7 | 104.5 | 99.5 | 28.5 | 84.6 |
| 9 | P2 | 100 | 90.3 | 83.3 | 97.0 | 94.0 | 97.8 | 93.7 |
| 10 | P2 | 100 | 86.7 | 30.7 | 99.0 | 102.0 | 96.8 | 85.9 |
| 11 | P1 | 100 | 66.1 | 0 | 99.5 | 97.0 | 72.5 | 72.5 |
| 12 | P2 | 100 | 92.7 | 93.0 | 101.0 | 95.0 | 73.7 | 92.6 |
| 13 | P2 | 100 | 86.7 | 90.4 | 99.5 | 95.0 | 100 | 95.3 |
| 14 | P2 | 100 | 101.2 | 93.3 | 106.0 | 91.0 | 102.1 | 98.9 |
| 15 | P3 | 100 | 77.0 | 83.0 | 102.5 | 97.0 | 110.5 | 95.0 |
| 16 | P1 | 90 | 68.5 | 18.9 | 100.5 | 98.0 | 70.0 | 74.3 |
| 17 | P1 | 100 | 91.5 | 107.8 | 99.0 | 99.0 | 89.7 | 97.8 |
| 18 | P1 | 100 | 89.1 | 100.4 | 99.0 | 96.5 | 90.2 | 95.9 |
| 19 | P3 | 100 | 95.2 | 68.5 | 101.0 | 95.5 | 102.8 | 93.8 |
| 20 | P3 | 100 | 92.7 | 70.0 | 96.0 | 96.5 | 97.9 | 92.2 |
| 21 | P1 | 100 | 100 | 86.3 | 99.0 | 98.0 | 97.5 | 96.8 |
| 22 | P1 | 100 | 100 | 93.3 | 106.5 | 97.5 | 95.3 | 98.8 |
| mean | | 99.5 | 88.8 | 79.1 | 100.5 | 96.8 | 79.0 | 90.6 |

| | SD | 2.1 | 9.3 | 27.8 | 2.6 | 2.9 | 31.0 | 7.7 |

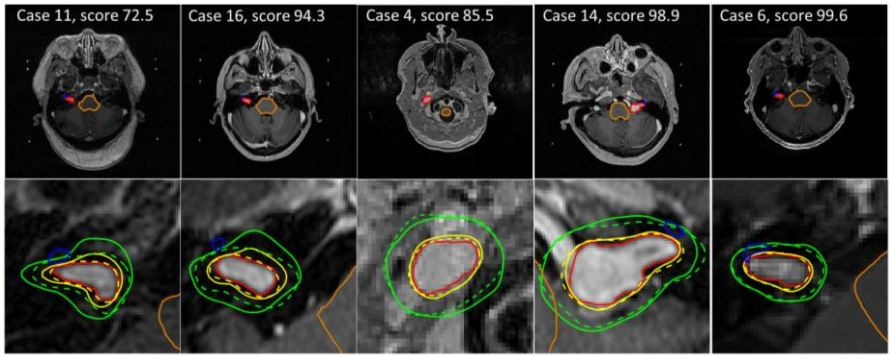

**Figure 1**. Isodose lines of the manual and golden plans for cases 11, 16, 4, 14, 6. They are arranged from left to right in ascending order of their manual plan score. Cases 11 and 16 have the lowest scores, and cases 14 and 6 that have the highest scores. The top row shows a slice of MR image for each case, overlaid with contours of target (red), brainstem (orange) and cochlea (blue). Cochlea is relatively far away from target in case 4 and is not shown. The bottom row shows the zoomed-in target region, with the isodose lines of the original manual plan (solid line) and the golden plan (dashed line). The isodose lines of prescription dose are shown in yellow and isodose lines of half of the prescription dose are shown in green.

We have also investigated plan quality variability for each planner. Large variability was found for planner P1, whose plan scores range from 72.5 (the lowest score in our study) to 98.8 (the second highest score). The average score for P1 is 87.8 with a standard deviation of 9.9. The plans generated by planner P2 are scored from 84.3 to 99.6, the mean score is 92.7 ± 5.2. The plans generated by planner P3 are scored from 92.2 to 95.0. Although he didn't produce a plan with a score higher than 95, he has the highest average score (93.7) and smallest standard deviation (1.4) among the three planners. These results not only demonstrate the inter-planner variability in GK planning, but also the intra-planner variability.

## 4. DISCUSSIONS AND CONCLUSIONS

There have been a few metrics proposed to evaluate a GK plan, such as coverage, selectivity and gradient index. However, an equivalent value of these metrics does not necessarily equate to the same quality level for different geometries, which is well demonstrated by the example of case 4 and case 6 shown in section 3. The manual plans of these two cases achieved the same coverage and selectivity, and similar gradient index. The relatively higher OAR doses and shorter BOT in case 6 seem to be reasonable due to its smaller distance between the target and the OARs and its smaller target volume. With these metric values, one may conclude that these two plans are of comparable quality level. In fact, the target of case 4 has a more spherical shape with a higher convexity than that of case 6. According to the nature of GK planning, case 4 is therefore less challenging for planning and one should expect better metric values than case 6. This impact of various patient geometries on GK planning needs to be considered in plan quality evaluation. Another confounding factor on plan quality evaluation is the different trade-offs adopted during treatment planning, which further challenges the plan evaluation and comparison among different cases. Therefore, in order to reduce these interferences on GK planning to objectively assess the plan quality and quality variation among different cases, in this study we assessed the plan quality for each specific case based on the deviation of the original manual plan from its corresponding golden plan. The golden plan was obtained by employing our recently developed MRL inverse planning approach. Its ability to yield a pareto optimal plan or a plan close to the pareto surface has been validated in our previous study[19]. Hence, we consider these golden plans generated by our computational optimization algorithms to be of a consistently high quality level.

To simplify our quality assessment, a scoring mechanism was proposed to calculate a single score for each case based on the quality disparity between the manual plan and the corresponding golden plan. This scoring mechanism was designed empirically, based on our observations and experiences during GK planning. We have found that although the exact score values depend on the values of the parameters involved in the scoring mechanism, the relative ranking of two plans at two different quality levels are insensitive to the parameters. It is critical to emphasize that the purpose of our scoring criterion is not to describe the best clinical plan, but to serve as a consistent, reasonable, and more importantly, relatively objective criterion to quantify the disparity between the manual plan and the golden plan for each case in our retrospective study.

There are two limitations of this retrospective study. One is the small amount of the patient cases investigated in this study, which is limited by the relatively low incidence of vestibular schwannoma compared to other brain tumors in general as well as the patient population in our institution. Another limitation is the considerable amount of human intervention involved. Since the golden plans were generated using our own planning algorithm outside of the commercial GK planning system LGP, these plans were put back to LGP for final dose calculation to ensure fair comparison. However, the current LGP system does not provide a data interface to import the shot locations, shapes and relative weights. The shot information must be manually entered into the system, which substantially increases the workload and is another limitation to the number of patient cases in our study. This issue can only be resolved by collaborating with Elekta to build a data interface for plans generated by other algorithms or systems, which will not only facilitate similar studies to this one, but also promote the development of external novel GK planning algorithms.

Although we only investigated 22 vestibular schwannoma cases, large variation of plan quality was observed among these cases. Variability was observed in both inter-planner and intra-planner treatment plan quality. This deserves high attention and merits further discussion and investigation on how to reduce the variation in GK treatment plan quality. Without knowing the best achievable plan quality for a given patient case, when to stop fine-tuning the plan and searching for potentially better quality can be a very subjective decision, relying on the experience and patience level of the planner at that moment. It also depends on the available time that can be devoted to that particular case. There are two potential ways to reduce the quality variation in GK planning. One is to develop a plan quality control system which can predict the best achievable plan quality for new patient cases based on their geometry. Machine learning or deep learning techniques can be employed to learn some prior knowledge from our previous cases to predict the plan quality for new cases[20-23]. Another way to reduce quality variation is to promote the development and clinical application of advanced planning algorithms for GK radiosurgery. Although there is a sequential inverse planning algorithm provided in the LGP system[24], it only optimizes the shots' sizes and durations at the predetermined isocenters and tends to be trapped in a local minimum, and thus can only provide a first approximation dose plan. Recently, a few inverse planning algorithms, including our MRL algorithm, have been developed that are able to optimize the isocenter locations and shots shapes and durations simultaneously[7,8,25,26]. However, an open data interface is much needed to facilitate the clinical validation and application of these external planning algorithms. Artificial intelligence may also be used to build and train a virtual GK planner with human-level intelligence[27,28].

**ACKNOWLEDGEMENTS**

This research is supported by Winship Cancer Institute #IRG-17-181-06 from the American Cancer Society.